\documentclass[superscriptaddress, twocolumn,amsmath, amssymb, aps, pra]{revtex4-2}

\newcommand{\ket}[1]{|#1\rangle} 
\newcommand{\bra}[1]{\langle#1|} 

\newcommand{\NaK}[0]{$^{23}$Na$^{39}$K }

\usepackage{graphicx}
\usepackage{cases}
\usepackage{multirow}
\usepackage{xcolor}
\usepackage{ulem}

\begin{document}

\title{Two-photon-assisted collisions in ultracold gases of polar molecules}

\author{Charbel Karam}
\affiliation{ Universit\'e Paris-Saclay, CNRS, Laboratoire Aimé Cotton, Orsay, 91400, France}
\affiliation{ Universit\'e Bourgogne Europe, CNRS, Laboratoire Interdisciplinaire Carnot de Bourgogne ICB UMR 6303, 21000 Dijon, France}
\author{Gohar Hovhannesyan}
\affiliation{ Universit\'e Paris-Saclay, CNRS, Laboratoire Aimé Cotton, Orsay, 91400, France}
\author{Romain Vexiau}
\affiliation{ Universit\'e Paris-Saclay, CNRS, Laboratoire Aimé Cotton, Orsay, 91400, France}
\author{Maxence Lepers}
\affiliation{ Universit\'e Bourgogne Europe, CNRS, Laboratoire Interdisciplinaire Carnot de Bourgogne ICB UMR 6303, 21000 Dijon, France}
\author{Nadia Bouloufa-Maafa}
\affiliation{ Universit\'e Paris-Saclay, CNRS, Laboratoire Aimé Cotton, Orsay, 91400, France}
\author{Olivier Dulieu}
\affiliation{ Universit\'e Paris-Saclay, CNRS, Laboratoire Aimé Cotton, Orsay, 91400, France}

\begin{abstract}
We present a theoretical formalism to treat the ultracold dynamics of a pair of colliding polar molecules submitted to two laser fields. We express the dressed Hamiltonian including the dipole-dipole interaction of the colliding molecular pair, both in their ground and electronic excited states, as well as their interaction with the two laser fields. We apply adiabatic elimination of the electronic excited state to reduce the size of the dressed-state basis in which the dressed Hamiltoninan is expressed. In an application, we investigate the feasibility of two-photon collisional shielding between two \NaK molecules, which could be favored by the Raman resonance condition suppressing unwanted spontaneous emission and photon scattering. We demonstrate the influence of the laser Rabi frequencies on the dynamics through the computation of elastic, inelastic, and reactive collision rates.
\end{abstract}

\maketitle

\section{Introduction}

Ultracold polar molecules represent a very promising platform for various domains of quantum sciences and technologies \cite{softley2023}, including quantum computation and simulation \cite{cornish2024} or ultracold chemistry \cite{bohn2023, karman2024}. Their great appeal comes from their permanent electric dipole moment (PEDM) ranging up to a few debyes, which induces anisotropic and long-range dipole-dipole interaction (DDI). The PEDMs of the molecules, as well as their rich internal structure, provide various knobs to precisely control them, for example, their optical cooling and trapping \cite{vexiau2017, langen2024}.

The field of ultracold molecules has recently experienced a ground-breaking achievement: the Bose-Einstein condensation of ground-state NaCs molecules \cite{bigagli2024}. This result followed the production of Fermi degenerate gases of KRb \cite{demarco2019} and NaK \cite{schindewolf2022}. In the three cases, it was necessary to overcome the instability of ultracold molecular samples due to trap losses \cite{zuchowski2010, demiranda2011, park2015, guo2016, gregory2019, voges2020} which are not yet fully understood \cite{mayle2013, croft2014, christianen2019b, gregory2020, liu2020, bause2021, jachymski2022, gersema2021}. This has been achieved by turning attractive long-range interactions between molecules into repulsive ones with electromagnetic fields mixing attractive and repulsive states: this defines the concept of collisional shielding, preventing pairs of molecules from getting close enough to each other. Two approaches have been successfully demonstrated, using static electric fields \cite{matsuda2020, li2021}, or microwave fields \cite{karman2018, lassabliere2018,karman2025, anderegg2021, bigagli2023, lin2023}, both inducing couplings between rotational levels of the ground-state molecules.

Another way of observing collisional shielding is to employ an optical field which induces coupling with an electronically excited state \cite{xie2022,xie2020}. This mechanism was observed in the 1990's with cold atoms in magneto-optical traps \cite{marcassa1994, bali1994, zilio1996}. However, spontaneous emission from the excited state during the collision has a detrimental effect on the shielding efficiency \cite{suominen1995}. Most importantly, photon scattering by individual molecules \cite{grimm2000} is expected to heat up an ultracold quantum gas.

To prevent photon scattering, we proposed  \cite{karam2023}, to create electromagnetically induced transparency (EIT) with two lasers coupling three molecular states in a {}``Lambda'' scheme \cite{fleischhauer2005}. We identified suitable molecular levels to perform such a two-photon transition, giving rise to repulsive potential energy curves (PECs) coupled by the photons to the attractive PECs of the colliding ground-state molecules, thus opening the possibility for an efficient two-photon shielding (2-OS).

To confirm this hypothesis, the static Hamiltonian for individual molecules must be extended to a dynamical Hamiltonian involving the scattering coordinates. In this article, we present the full formalism and methodology necessary to describe two-photon-assisted collisions between ultracold polar molecules, assuming that the collisional dynamics essentially takes place at a large intermolecular distance, where dipole-dipole and van der Waals interactions are dominant. Starting from the static EIT condition with a ``Lambda'' scheme for the individual molecule involving an electronic excited state, we introduce the scattering coordinate which results in a ''dynamical'' model with a double ``Lambda'' scheme with five levels. It differs from the conventional five-level-static scheme \cite{li2011b} in the fact that the resonance condition varies with the scattering coordinate \cite{karam2023}. Our approach relies on the adiabatic elimination of the electronic excited state, thus reducing the computational cost while maintaining the precision of the results. We apply this formalism to the interactions between two \NaK molecules, extracting intermolecular PECs dressed by the two photons. We apply our formalism to calculate rate coefficients of elastic, inelastic, and reactive collisions between molecules, in order to determine the feasibility of the 2-OS mechanism in experimentally realistic conditions.

The structure of the paper is as follows. Section \ref{sec:2Phot} develops the formalism of two-photon transitions in interacting molecules. We begin by presenting the general Hamiltonian of the system, followed by a detailed construction of the basis set, accounting for the system’s symmetries. We also derive the selection rules that govern the transitions within the system. 
In Section \ref{sec:RelBasSt}, we focus on the relevant basis states, building a coupling scheme that enables the construction of the matrix Hamiltonian in the noninteracting region, which we later extend to the full $R$-dependent Hamiltonian. This section also includes the potential energy curves for the two \NaK molecules exposed to the two lasers, providing insights into how the optical fields influence the molecular interaction.
Section \ref{sec:AdiabElim} introduces the adiabatic elimination approximation and explores its validity when applied to the potential energy curves. We analyze how this approximation simplifies the calculations and reduces the complexity of the problem while preserving the physical accuracy of the model. 
Then, Section \ref{sec:2OS} applies the formalism developed in the previous sections to the 2-OS scheme \cite{karam2024}. Here, we solve the coupled-channel equations which allow us to compute and analyze the elastic, inelastic, and reactive collision rates. This practical application demonstrates the utility of our formalism and highlights its potential for controlling molecular interactions in ultracold systems. Finally, Section \ref{sec:ccl} contains conclusions and prospects.

\section{Two-Photon Transitions in Interacting Molecules}
\label{sec:2Phot}

In this section, we give a general description of the long-range interactions between two molecules in the presence of two laser fields. We consider two diatomic heteronuclear molecules described by their electronic, vibrational, and rotational quantum numbers. We use the projection of the orbital angular momentum $\Lambda$ and of the total angular momentum $\Omega$ on the molecular axis, as well as the spin quantum number $S$. Our presentation is general and we consider alkali-metal diatomic molecules in their electronic ground state $X^1\Sigma^+$ ($S = \Lambda = \Omega = 0$), and in the long-lived electronic excited state $b^3\Pi_{0^+}$ ($S=1$, $\Lambda=\pm 1$, $\Omega=0$). In these two states, a given rotational level $j$ is characterized by a single parity $p=(-1)^j$. For the interaction with light, we focus on a {}``Lambda'' configuration of the Raman beams.

\subsection{General Hamiltonian}

The Hamiltonian $\hat{H}_{\text{int}}(R)$ for two interacting molecules, under the influence of two laser fields, is expressed as
\begin{equation} \label{H_int}
\begin{split}
  \hat{H}_{\text{int}}(R) = \sum_{i=1}^2 \hat{H}_0(i) +
   & \frac{\hat{L}^2}{2 \mu R^2} + \hat{V}_{\text{int}}(R) \\
   & + \sum_{\alpha=1}^2 \Big( \hat{H}_{L_\alpha} 
   + \sum_{i=1}^2 \hat{H}^{(\alpha)}_{ac}(i) \Big),
\end{split}
\end{equation}
with $i=1,2$ and $\alpha=1,2$ numbering the molecules and the lasers, respectively.  The Hamiltonian $\hat{H}_0(i)$ of the bare molecule $i$ includes contributions from the electronic ($E_e$), vibrational ($E_{\text{vib}}$), and rotational energies. Its matrix elements can be explicitly written as $\langle \hat{H}_0(i) \rangle = E_e(i) + E_{\text{vib}}(i) + B(i) \hat{j_i}^2$, where $B(i)$ is the rotational constant, and $\hat{j}_i$ the angular momentum of the molecule $i$ with quantum number $j_i$. The centrifugal term $\hat{L}^2/(2 \mu R^2)$ involves the angular momentum $\hat{L}$ (with the quantum number $\ell$ generally termed the partial wave) of the rotation of the intermolecular axis where $\mu$ the reduced mass of the two molecules. The term $\hat{V}_{\text{int}}(R)$ represents the molecule-molecule interaction potential, and $\hat{H}_{L_\alpha}$ the energy of laser $\alpha$. It can be written as \cite{cohen-tannoudji1998}
\begin{equation}
\hat{H}_{L_\alpha} = \hbar \omega_{L_\alpha} \big(\hat{a}_\alpha^{\dagger}\hat{a}_\alpha + \frac{1}{2}\big),
\end{equation}
where $\hbar \omega_{L_\alpha}$ is the energy of a photon in the laser $\alpha$ and $\hat{a}_\alpha$ and $\hat{a}_\alpha^{\dagger}$ are the annihilation and creation operators of the laser field $L_\alpha$, respectively. The laser fields are also called the Stokes and anti-Stokes beams. The light-matter Hamiltonian $\hat{H}^{(\alpha)}_{ac}(i)$ for laser $L_\alpha$ with molecule ($i$) is 

\begin{equation}
  \hat{H}^{(\alpha)}_{ac} (i) = g_\alpha \big({\cal \hat{S}}_{\alpha}^+(i) + {\cal \hat{S}}_{\alpha}^-(i)\big) (\hat{a}_\alpha^{\dagger} + \hat{a}_\alpha),
  \label{eq:Hlight}
\end{equation}
where the excitation operators ${\cal \hat{S}}_{\alpha}^{\pm}$ excite a molecule from a state $\ket{g}$ to a state $\ket{e}$. The coupling coefficients $g_\alpha$ are defined by
\begin{equation}
  {\cal \hat{S}}_{\alpha}^+ = {\cal \hat{S}}_{\alpha}^-{^{\dagger}} 
    = \ket{g} \bra{e} \quad \text{and} \quad 
  g_\alpha = -\vec{\epsilon}_{L_\alpha} \cdot \vec{d}_{ge} 
  \sqrt{\frac{\hbar \omega_{L_\alpha}}{2 \epsilon_0 V}},
\end{equation}
where $\vec{d}_{ge}$ is the transition electric dipole moment (TEDM) of the transition $\ket{g} \leftrightarrow \ket{e}$, $\vec{\epsilon}_{L_\alpha}$ is the polarization vector of laser $L_\alpha$, and $V$ is the quantization volume of the laser field. Here, $\vec{\epsilon}_{L_\alpha}$ and $\vec{d}_{ge}$ are assumed real. 

In the case of two neutral dipolar molecules, the main term of the intermolecular potential comes from the dipole-dipole interaction (DDI). We write the interaction potential $\hat{V}_{\text{int}}(R)$ in the space-fixed (SF) frame with spherical coordinates ($R, \Theta, \varPhi)$) of the intermolecular axis as \cite{lepers2018}
\begin{equation}\label{multipol_SF_spherical}
  \begin{split}
    &V_{\text{int}}(R,\Theta,\varPhi)= -\frac{1}{4 \pi \epsilon_0} \sqrt{\frac{4 \pi}{5}} \frac{1}{R^3} \\
    & \times \sum_{m=-2}^{2} {Y_2^m}^*(\Theta, \varPhi) \sqrt{(2 +m)! (2 -m)!}    \\
    & \times \sum_{q_1=-1}^{1} \sum_{q_2=-1}^{1} \frac{Q_{1 q_1}(1)Q_{1 q_2}(2)}{\sqrt{(1+q_1)!(1-q_1)!(1+q_2)!(1-q_2)!}},
  \end{split} 
\end{equation}
with $\epsilon_0$ is the vacuum permeability, ${Y_2^m}(\Theta, \varPhi)$ are the spherical harmonics acting on the angular coordinates in the SF frame, and $Q_{1 q_1}(1)$ and $Q_{1 q_2}(2)$ are the SF dipole operators of the first and second molecule, respectively, corresponding to the tensorial form of $\vec{d}$ introduced above.


\subsection{Basis set and symmetries}

The Hamiltonian \eqref{H_int} is expressed in a dressed-state picture. The specificity of our system is that the dressed {}``particle'' is the collisional complex formed by the two molecules, whose basis states are noted as $|1,2\rangle$. Moreover, the dressing is made by two laser fields with photon numbers $N_1$ and $N_2$, generating the so-called Fock states $\ket{N_1,N_2}$. The two-molecule, two-field-dressed basis states can thus be written as the tensor product $\ket{1,2} \otimes \ket{N_1,N_2}$.

\subsubsection{Uncoupled \textit{versus} coupled basis}

There are two different choices for coupling the angular momenta $\vec{j}_1$, $\vec{j}_2$ and $\vec{L}$. A first choice is the uncoupled dressed basis
\begin{equation}\label{uncoupled_basis+fock}
\begin{split}
  & \ket{n} = |\beta_1 j_1 m_1 p_1, \beta_2 j_2 m_2 p_2,
   \ell m_\ell\rangle \otimes \ket{N_1,N_2},
\end{split}
\end{equation}
where $\beta_i$ the quantum numbers for the vibronic state of molecule $(i)$, $m_i$ and $m_\ell$ the projection of $\vec{j}_i$ and $\vec{L}$ on the quantization axis, and $p_i$ the parity. 

The other choice is the fully coupled dressed basis where $\Vec{j}_1$ and $\Vec{j}_2$ are first coupled to produce $\Vec{j}_{12}$ with quantum number $j_{12}$, which is then coupled to $\vec{L}$ to form the total angular momentum $\vec{J}$ with a projection $M$ on the quantization axis,
\begin{equation}\label{coupled_basis+fock}
   \ket{n} = |\beta_1 j_1 p_1, {\beta_2 j_2 p_2}, j_{12} \ell, J M\rangle \otimes \ket{N_1,N_2}. 
\end{equation}
The relation between the two bases is the following (omitting for simplicity the tensorial product with the Fock states which are unaffected),
\begin{equation}\label{coupled_to_uncoupled}
 \begin{split}
  |\beta_1 j_1& p_1, \beta_2 j_2 p_2, j_{12} \ell, J M\rangle
    \\
   = \sum_{m_{12}, m_{\ell}} C^{JM}_{j_{12} m_{12} \ell m_\ell} 
   & \: |\beta_1 j_1 p_1, \beta_2 j_2 p_2, j_{12} m_{12},
   \ell m_\ell \rangle \\
   = \sum_{\substack{{m_1},{m_2}\\
           {m_{12}, m_{\ell}}}} C^{JM}_{j_{12} m_{12} \ell m_{\ell}} 
    & C^{j_{12} m_{12}}_{{j_1 m_1} {j_2 m_2}} \\
    \times |\beta_1 & j_1 m_1 p_1, \beta_2 j_2 m_2 p_2, 
    \ell m_{\ell} \rangle ,
 \end{split}
\end{equation}
involving the Clebsch-Gordon coefficients $C_{\beta b \gamma c}^{\alpha a }$. We have used the coupled basis in our calculations, whereas the uncoupled basis will be used to interpret our results. 

In the absence of external fields, $J$ and $M$ are good quantum numbers. If the fields are polarized along the $z$ axis, $M$ remains a good quantum number. In practice, the quantum numbers $j_1$, $j_2$ and $\ell$ are taken up to maximum values $j_{1_{\mathrm{max}}}$, $j_{2_{\mathrm{max}}}$ and $\ell_{\mathrm{max}}$, ensuring convergence of the computed quantities. It is also the case for the total angular momentum $J$ in the coupled basis; and when
$J_{\mathrm{max}} = j_{1_{\mathrm{max}}} + j_{2_{\mathrm{max}}} + \ell_{\mathrm{max}}$, the two basis sets become equivalent under the condition $m_1+m_2+m_{\ell}=M$.

\subsubsection{Additional symmetries}

An important aspect of the present study is the symmetries that can be utilized to further structure the Hamiltonian matrix, leading to more efficient calculations \cite{li2019}. For two identical interacting particles, we consider three primary symmetries: inversion $\hat{E}$, permutation $\hat{P}_{12}$, and reflection $\hat{\sigma}_{xz}$ (the latter only applies when $M=0$). We here show the effect of each of these symmetry operations on both the uncoupled and coupled basis. Since these operations do not act on the Fock states, we omit $\ket{N_1,N_2}$ for simplicity.

The inversion symmetry is defined by
\begin{equation}\label{parity}
\begin{cases}
\!\begin{aligned}
  & \hat{E} \:| \:{\beta_1 j_1 m_1 p_1}, {\beta_2 j_2 m_2 p_2}, \ell m_{\ell} \rangle \\
  & = p_1 p_2 (-1)^{\ell} \: |\:{\beta_1 j_1 m_1 p_1}, {\beta_2 j_2 m_2 p_2}, \ell m_{\ell} \rangle
  \\[2.5ex]
  & \hat{E} \:| \:{\beta_1 j_1 p_1}, {\beta_2 j_2 p_2}, \: j_{12} \ell, \: J M\rangle \\
  & = p_1 p_2 (-1)^{\ell} \: |\:{\beta_1 j_1 p_1}, { \beta_2 j_2 p_2}, \: j_{12} \ell, J M\rangle,
\end{aligned}
\end{cases}    
\end{equation}
indicating that whatever basis is used, a state of the complex has a well-defined total parity $P=p_1 p_2 (-1)^{\ell}$.

The permutation symmetry is defined by
\begin{equation}\label{permutation}
\begin{cases}
\!\begin{aligned}
    &\hat{P}_{12} \:| \:{\beta_1 j_1 m_1 p_1} ,{\beta_2 j_2 m_2 p_2} , \ell m_{\ell} \: \rangle \: \\
    &= (-1)^{\ell} \:|\: {\beta_2 j_2 m_2 p_2}, {\beta_1 j_1 m_1 p_1}, \ell m_{\ell} \:\rangle
  \\[2.5ex]
    &\hat{P}_{12} \:| \:{\beta_1 j_1 p_1} ,{\beta_2 j_2 p_2}, \: j_{12} \ell, J M\: \rangle, \\
    &= (-1)^{{j_1}+{j_2}-j_{12}+\ell} \:|\: {\beta_2 j_2 p_2},{\beta_1 j_1 p_1},\:\: j_{12} \ell, J M\rangle .
\end{aligned}
\end{cases}    
\end{equation}
The reflection operation $\hat{\sigma}_{xz}$ with respect to the plane $xz$ is only relevant when $M=m_1+m_2+m_{\ell}=0$. It is equivalent to an inversion followed by a rotation of $\pi$ radians around the $y$ axis, which gives
\begin{equation}\label{reflection}
\begin{cases}
\!\begin{aligned}
   & \hat{\sigma}_{xz}\:| \: {\beta_1 j_1 m_1 p_1},\: {\beta_2 j_2 m_2 p_2}, \: \ell m_{\ell}\rangle \\
   &= p_1 p_2 (-1)^{j_1+j_2}
   \:| {\beta_1 j_1(-m_1)p_1},{\beta_2 j_2(-m_2)p_2}, \ell(-m_{\ell}) \rangle.
  \\[2.5ex]
   & \hat{\sigma}_{xz}\:| \: {\beta_1 j_1 p_1} ,{\beta_2 j_2 p_2}, \: j_{12} \ell J 0\rangle, \\
   &= p_1 p_2 (-1)^{\ell+J} \:| \: {\beta_1 j_1 p_1}, {\beta_2 j_2 p_2}, \: j_{12} \ell J 0\rangle.
\end{aligned}
\end{cases}    
\end{equation}
A given coupled basis state $| \: {\beta_1, j_1 p_1} ,{\beta_2 j_2 p_2}, \: j_{12} \ell, J 0\rangle$ has a well-defined even or odd character $p_1 p_2 (-1)^{\ell+J}$ under reflection. 
As the parity of $^1\Sigma$ states or $\Omega=0$ states of individual molecules is equal to $(-1)^{j_i}$, we have $p_1 p_2 (-1)^{j_1+j_2} = 1$ in these special cases.

In the case of identical particles, it is useful to use the symmetrized basis (identified with the brackets $[\:]$) 
\begin{equation}\label{symmetrized_basis}
\begin{cases}
&\begin{split}
       & |[\beta_1 j_1 m_1 p_1, {\beta_2 j_2 m_2 p_2}], \ell m_{\ell} ; {\eta} \rangle \\
      &  = \frac{1}{\sqrt{2(1+ \delta_{\beta_1,\beta_2} \delta_{j_1,j_2} \delta_{m_1,m_2} \delta_{p_1,p_2}})} \\
      &  \times \big( |\beta_1 j_1 m_1 p_1, {\beta_2 j_2 m_2 p_2}, \ell m_{\ell}\rangle\\
       &+ \eta \varepsilon  |{\beta_2 j_2 p_2}, \beta_1 j_1 p_1,  \ell m_{\ell} \rangle \big)
    \end{split}\\
    &\\
 &   \begin{split}
       & |[\beta_1 j_1 p_1, {\beta_2 j_2 p_2}] j_{12} \ell J M; {\eta} \rangle \\
      &  = \frac{1}{\sqrt{2(1+ \delta_{\beta_1,\beta_2} \delta_{j_1,j_2} \delta_{p_1,p_2}})} \\
      &  \times \big( |\beta_1 j_1 p_1, {\beta_2 j_2 p_2}, j_{12} \ell J M\rangle\\
       &+ \eta \varepsilon' |{\beta_2 j_2 p_2}, \beta_1 j_1 p_1,  j_{12} \ell J M\rangle \big)
    \end{split}
 \end{cases}
\end{equation}
with $\varepsilon = (-1)^{\ell}, \varepsilon' = (-1)^{j_1+j_2- j_{12}+\ell}$, and $\eta = + 1$ for identical bosons and $\eta =- 1$ for identical fermions. From now on, since we work with identical bosons, we will use the symmetrized basis with $\eta=+1$ and omit the symbol $\eta$ for simplicity.

\subsection{Selection rules}

The selection rules associated with the DDI can be obtained by calculating the matrix elements of the operator $V_{\text{int}}(R,\Theta,\varPhi)$ (Eq. \eqref{multipol_SF_spherical}), expressed in the uncoupled or coupled basis above. The DDI operator couples states within the same Fock state ($\Delta N_1 = \Delta N_2 = 0$), and acts on the quantum numbers of both molecules at the same time, changing the parity and rotation according to \cite{lepers2018}: $|\Delta j_1| = 1$, $p_{1}p'_{1} = -1$, $|\Delta j_2| = 1$, $p_{2}p'_{2} = -1$. Note that $\Delta j_i = 0$ is not possible since we consider $\Sigma^+$ and $\Pi_{0^+}$ electronic states. Regarding the quantum numbers of the complex, we have $|\Delta \ell| = 0,2$, $\Delta J=0$ and $\Delta M=0$. These selection rules are summarized in Table \ref{tab:Selection_rules_DDI}.

\begin{table}[h!]
  \begin{center}
  \caption{Selection rules of the DDI (Eq. \eqref{multipol_SF_spherical}), written for the symmetrized fully-coupled basis states.}
  \label{tab:Selection_rules_DDI}
  \begin{tabular}{cc}
    \hline \hline
    Quantum number & Selection rule \\
    \hline\hline
      \multirow{2}{*}{$[p_1,p_2]$} &
        $[\pm,\pm] \leftrightarrow [\mp,\mp]$  \\
      & $[\pm,\mp] \leftrightarrow [\mp,\pm]$ \\ [0.3ex]
      $[\Delta j_1,\Delta j_2]$ & $[\pm 1,\pm 1]$ or $[\pm 1,\mp 1]$  \\[0.3ex]
      $\Delta j_{12}$ & $0^{(*)}$ or $\pm 1$ or $\pm 2$  \\[0.3ex]
      $\Delta L$ & $0^{(*)}$ or $\pm 2$\\[0.3ex]
      $\Delta J$ & 0  \\[0.3ex]
      $\Delta M$ & 0 \\[0.3ex]
      Parity & $\pm \leftrightarrow \pm$  \\[0.3ex]
      Reflection & $\pm \leftrightarrow \pm$  \\[0.3ex]
      Permutation & $\pm \leftrightarrow \pm$  \\[0.3ex]
    \hline\hline
    \multicolumn{2}{l}{$^*$ $\Delta X=0$ except $0 \leftrightarrow 0$}
    \\
  \end{tabular}
  \end{center}
\end{table}

The light coupling operator \eqref{eq:Hlight} acts on the quantum numbers of one molecule, while those of the other molecule are unchanged. We express the matrix elements of $\sum_{i=1}^2 \hat{H}^{(\alpha)}_{ac}(i)$ on the basis of the individual molecule $\ket{\beta'_i, j'_i, m'_i,p'_i}$ in terms of the Rabi frequency $\Omega_\alpha$ of $L_{\alpha}$ as
\begin{equation}\label{light_coupling_Rabi}
  \begin{split}
  \bra{\beta'_i j'_i m'_i p'_i}
    \hat{H}^{(\alpha)}_{ac}(i) \ket{\beta_i j_i m_i p_i} \\
  = C^{j'_i, m'_i}_{j_i, m_i, 1, q}
    C^{j'_i, \Omega'_i}_{j_i, \Omega_i, 1, \Omega'_i-\Omega_i} 
    \sqrt{\frac{2j_i + 1}{2j'_i + 1}} \Omega_{\alpha,i} \,,
  \end{split}
\end{equation}
where $\Omega_{\alpha,i}$ is the vibronic Rabi frequency -- it should not be mixed up with the projection $\Omega_i$ of the total angular momentum on the molecular axis --, $q=0$ for linear polarization ($\pi$) and $q=\pm 1$ for circular ($\sigma^{\pm}$) polarization. The vibronic Rabi frequency is a function of the matrix element of the dipole moment $d_\mu(i)$ expressed in the molecular frame and of the amplitude of the electric field $\alpha$, that is, $\Omega_{\alpha,i} = \bra{\beta'} d_{\Omega'-\Omega}(i) \ket{\beta} \times \mathcal{E}_\alpha$.

The selection rules associated with Eq.~\eqref{light_coupling_Rabi} are given in Table \ref{tab:selection_rules_1photon} for single-molecule quantum numbers for $\sigma^\pm$ and $\pi$ light polarizations, which address different projection quantum numbers $\Delta m = \pm 1$ and 0, respectively. The Clebsch-Gordan coefficient $C^{j'_i, \Omega'_i}_{j_i, \Omega_i, 1, \Omega'_i-\Omega_i}$ in Eq.~\eqref{light_coupling_Rabi} prevents $j'_i = j_i$ for $\Omega'_i = \Omega_i$, or $\Lambda'_i = \Lambda_i$ when Hund's case (a) applies.

\begin{table} [h!]
  \begin{center}
  \caption{Selection rules of the electric dipole-dipole interaction, the one-photon and two-photon electric-dipole optical transitions, written for the symmetrized  fully-coupled basis states.}
  \label{tab:selection_rules_1photon}
  \begin{tabular}{ccc}
    \hline \hline
    Quantum & One-photon  & One-photon  \\
    numbers   & $\pi$ transition & $\sigma^{\pm}$ transition \\
    \hline\hline
      $\Delta S$ & $0$ & $0$ \\ [0.3ex]
      $\Delta \Lambda$ & $0, \pm 1$ & $0, \pm 1$ \\[0.3ex]
      $\Delta \Omega$  & $0, \pm 1$ & $0, \pm 1$ \\[0.3ex]
      $\Delta j^{\text{a}}$ & $0, \pm 1^{\text{b}}$ & $0, \pm 1$ \\[0.3ex]
      $\Delta m$ & $0$ & $\pm 1 $ \\[0.3ex]
      Parity & $\pm \leftrightarrow \mp$ & $\pm \leftrightarrow \mp$ \\[0.3ex]
     \hline \hline
     \multicolumn{3}{l}{ $^{\text{a}}$ In the case $\Lambda =0 \rightarrow \Lambda =0$ or $\Omega =0 \rightarrow \Omega =0$,}\\
     \multicolumn{3}{l}{only $\Delta j=\pm 1$}\\
     \multicolumn{3}{l}{ $^{\text{b}}$ $\Delta j = \pm 1$ for $m=m'=0$}\\
    \end{tabular}
  \end{center}
\end{table}

Regarding the basis states of the colliding molecules, the light coupling does not act on $\ell$ and $m_\ell$, since $\hat{L^2}$ commutes with $\hat{H}^{(\alpha)}_{ac}(i)$. In the fully coupled basis, one has $\Delta j_{12} = 0, \pm 1$, as well as $\Delta J = 0, \pm 1$ when $\Delta M = \pm 1$ ($\sigma^\pm$ polarizations) and $\Delta J = \pm 1$ when $\Delta M = 0$ ($\pi$ polarizations). The total parity of the complex is modified ($\pm \leftrightarrow \mp$), while the reflection and permutation symmetries are unchanged \cite{li2019}.

The selection rules for two-photon transitions are obtained by applying twice the selection rules of one-photon transitions given in Table \ref{tab:selection_rules_1photon}. In particular, one obtains $0 \le |\Delta j_i| \le 2$, $\Delta m_i = q_1-q_2$, where $q_1$ and $q_2$ define the polarization of the Stokes and anti-Stokes beams, and the one-molecule parity is unchanged ($\pm \leftrightarrow \pm$). Regarding the complex, the partial wave is still unchanged $(\Delta \ell = 0)$, while $0 \le |\Delta J| \le 2$ and $\Delta M = q_1-q_2$.

\section{Dynamic multi-level dressed model for two interacting molecules}
\label{sec:RelBasSt}

Following our previous work \cite{karam2023}, we consider a bosonic \NaK molecule exposed to two lasers with frequencies tuned to the two-photon transition
\begin{equation}
   \begin{split}
    \ket{X{}^1 \Sigma^+, v_X=0, j_X=0} 
    \rightarrow \ket{b{}^3 \Pi, v_b=0, j_b=1} \\
    \rightarrow \ket{X{}^1 \Sigma^+, v_X=0, j_X=2}
   \end{split} 
\end{equation}
These three levels represent the static {}``Lambda'' scheme for the isolated molecule, associated, respectively, with the Fock states $\ket{N_1,N_2}$, $\ket{N_1-1,N_2}$, $\ket{N_1-1,N_2+1}$. The two-photon resonance is reached when $\hbar ( \omega_{L_1} - \omega_{L_2} ) = 6B_X$, with $B_X$ the rotational constant of $X{}^1 \Sigma^+ (v_X=0)$. Note that the upper level of this ``Lambda'' scheme could belong to any $\Sigma$ or $\Pi$ electronic state with $\Omega = 0, \pm 1$ having a sufficient TEDM with $X{}^1 \Sigma^+$.

In what follows, we show how to transfer this choice of states to the case of two noninteracting molecules. 

\subsection{Asymptotic study: Non-Interacting Molecules}

We first examine the two molecules at infinite separation ($R \to \infty$), where they do not interact \cite{napolitano1997, napolitano1998}. In the absence of an electromagnetic field, each molecule rotates freely with a rotational energy $ B(i) j_i(j_i+1) $. When two lasers are applied to the molecules, the interaction Hamiltonian (Eq. \ref{H_int}) reduces to the asymptotic Hamiltonian $H_{\infty}$
\begin{equation}\label{infty_Hamiltonian}
\hat{H}_{\infty} = \sum_{i=1}^2 \hat{H}_0(i) + \sum_{\alpha=1}^2 \big( \hat{H}_{L_\alpha} + \sum_{i=1}^2 \hat{H}^{(\alpha)}_{ac}(i) \big),
\end{equation}
%

For pedagogical reasons, we use the symmetrized and uncoupled dressed basis set $\ket{ [\beta_1j_1m_1p_1, \beta_2j_2m_2p_2}] \otimes \ket{N_1,N_2}$, combining molecular and Fock states, and disregarding partial waves since $R \to \infty$. We use below the short notation for the vibronic states $X \equiv \ket{X{}^1 \Sigma^+, v_X=0}$ with energy $E_X=0$ for $j_X=0$, and $b \equiv \ket{b{}^3 \Pi, v_b=0}$ with energy $E_b$ for $j_b=0$.

The two ground-state molecules undergoing a two-photon transition at infinity are now described by the following five states (instead of three states for a single molecule)
\begin{equation}
\begin{cases}
    \ket{g_1} = \ket{ [X,0,m_1,+1, X,0,m_2,+1]} \otimes \ket{0,0}, \\
    \ket{g_2} = \ket{ [X,0,m_1,+1, X,2,m_2,+1]} \otimes \ket{-1,+1}, \\
    \ket{g_3} = \ket{ [X,2,m_1,+1, X,2,m_2,+1]} \otimes \ket{-2,+2}, \\
    \ket{e_1} = \ket{ [X,0,m_1,+1, b,1,m_2,-1]} \otimes \ket{-1,0},  \\
    \ket{e_2} = \ket{ [X,2,m_1,+1, b,1,m_2,-1]} \otimes \ket{-2,+1}.
  \end{cases}
  \label{eq:bas5st}
\end{equation}
We keep the study general for any value of the projections $m_1$ and $m_2$ depending on which values are allowed by the selection rules. Without loss of generality, we consider $N_1 = N_2 = 0$ in the initial state $g_1$, since non-zero values of $(N_1,N_2)$ only give a global shift in energy.
\begin{figure}[h!]
    \centering
    \includegraphics[scale=0.38]{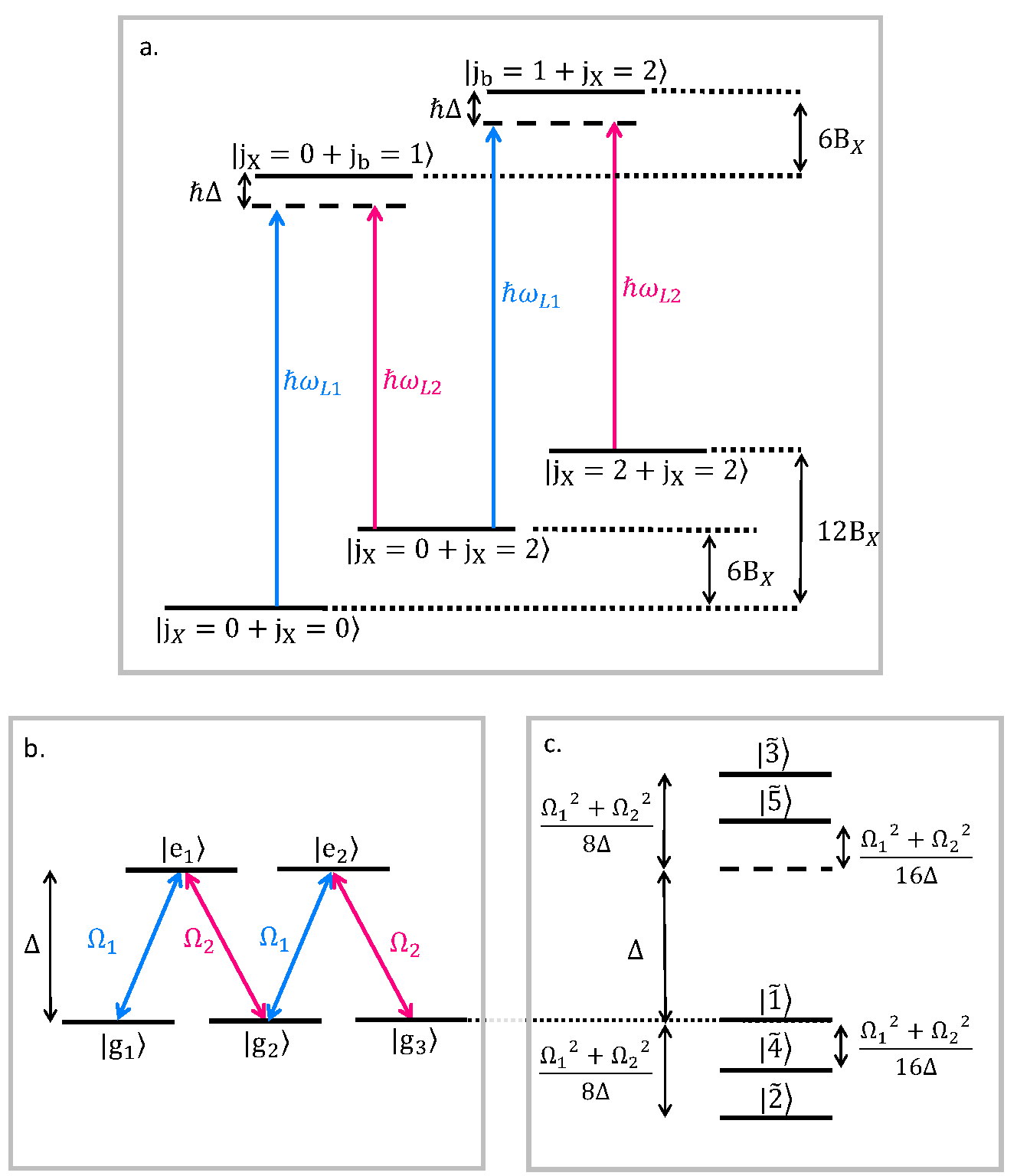}
    \caption{Five-level system at Raman resonance ($\delta\omega=6B_X$): On the left, energy levels in the rotating wave frame with couplings from lasers $L_1$ (blue) and $L_2$ (pink). On the right, dressed energy levels after diagonalizing the asymptotic Hamiltonian \eqref{eq:H5States-2}. Horizontal dashed lines mark the energy origin in both panels.}
    \label{fig:5-level-system}
\end{figure}

As illustrated in Fig. \ref{fig:5-level-system}, in the specific case where both lasers are equally detuned by respect to the intermediate states, the first (Stokes) laser couples $\ket{g_1}$ to $\ket{e_1}$ and $\ket{g_2}$ to $\ket{e_2}$, and the second (anti-Stokes) laser couples $\ket{g_2}$ to $\ket{e_1}$ and $\ket{g_3}$ to $\ket{e_2}$. We express the matrix of the Hamiltonian in Eq.  \ref{infty_Hamiltonian} in the basis of the five states $\{\ket{g_1}, \ket{g_2}, \ket{g_3}, \ket{e_1}, \ket{e_2}\}$ as
\begin{widetext}
\begin{equation}
  \hat{H}_{\infty} = \begin{pmatrix} 
    0 & 0 & 0 & {\hbar {\Omega_1}}/{2\sqrt{2}} & 0 \\
    0 & 6B_X - \hbar (\omega_{L_1} - \omega_{L_2}) & 0 & \hbar \Omega_2/4 & {\hbar \Omega_1}/{4} \\
    0 & 0 & 12B_X - 2 \hbar (\omega_{L_1} - \omega_{L_2}) & 0 & \hbar \Omega_2/2\sqrt{2} \\
    {\hbar \Omega_1}/{2\sqrt{2}} & {\hbar \Omega_2}/{4} & 0 & E_b + 2B_b - \hbar \omega_{L_1} & 0 \\
    0 & {\hbar \Omega_1}/{4} & {\hbar \Omega_2}/{2\sqrt{2}} & 0 & E_b + 2B_b + 6B_X - \hbar (2\omega_{L_1} - \omega_{L_2})
  \end{pmatrix}.
  \label{eq:H5States-1}
\end{equation}
where $(\Omega_1,\Omega_2)$ are the Rabi frequencies of the two beams. Unlike the vibronic Rabi frequency $\Omega_{\alpha,i}$ introduced in Eq.~\eqref{light_coupling_Rabi}, $\Omega_1$ and $\Omega_2$ depend on the laser polarizations $q_1$ and $q_2$. Equation \eqref{eq:H5States-1} can be rewritten as a function of the detuning defined with respect to the energy of the transition between $\ket{g_1}$ and $\ket{e_1}$ as $\hbar \Delta = E_b + 2B_b - \hbar \omega_{L_1}$ and the two-photon detuning $\delta\omega = \omega_{L_1} - \omega_{L_2}$,
\begin{equation}
  \hat{H}_{\infty} = \begin{pmatrix} 
    0 & 0 & 0 & {\hbar \Omega_1}/{2\sqrt{2}} & 0 \\
    0 & 6B_X -  \hbar \delta\omega & 0 & \hbar \Omega_2/4 & {\hbar \Omega_1}/{4} \\
    0 & 0 & 12B_X - 2 \hbar \delta\omega & 0 & \hbar \Omega_2/2\sqrt{2} \\
    {\hbar \Omega_1}/{2\sqrt{2}} & {\hbar \Omega_2}/{4} & 0 & \hbar \Delta & 0 \\
    0 & {\hbar \Omega_1}/{4} & {\hbar \Omega_2}/{2\sqrt{2}} & 0 & \hbar \Delta + 6B_X - \hbar \delta\omega
  \end{pmatrix},
  \label{eq:H5States-2}
\end{equation}
\end{widetext}
In this context, we define the Raman resonance as $\hbar \delta\omega = 6B_X$. Note that a blue (red) detuning corresponds to a negative (positive) value of $\Delta$. Note also that the $1/2\sqrt{2}$ and $1/4$ prefactors in the off-diagonal terms come from the normalizing factor of the symmetrized basis states which take into account the permutation of the molecules, and which appear in the explicit expressions of the states $\{ \ket{g_1}, \ket{g_2}, \ket{g_3}, \ket{e_1}, \ket{e_2}\}$ using Eq. (\ref{symmetrized_basis}).

At the Raman resonance, one of the eigenstates in Eq. \ref{eq:H5States-2} is a dark state, composed only of electronic ground states $\ket{g_1}, \ket{g_2}$ and $\ket{g_3}$. Unlike the other eigenstates, the energy of this dark state is not shifted by light (Fig. \ref{fig:5-level-system}c). In the weak field regime ($\Omega_1,\Omega_2 \ll \Delta$), the dressed energies of $\hat{H}_{\infty}$ are
\begin{equation}
\begin{cases}
  E_1 = 0, \\
  \displaystyle E_2 \approx - \frac{\Omega_1^2 + \Omega_2^2}{8 \Delta},\\
  \displaystyle E_3 \approx \Delta + \frac{\Omega_1^2 + \Omega_2^2}{8 \Delta}, \\
  \displaystyle E_4 \approx -\frac{\Omega_1^2 + \Omega_2^2}{16 \Delta}, \\
  \displaystyle E_5 \approx \Delta + \frac{\Omega_1^2 + \Omega_2^2}{16 \Delta},
\end{cases}
\end{equation}
with dressed eigenvectors
\begin{widetext}
\begin{equation} \label{dressed_states_5lvl}
\begin{cases}
  \displaystyle \ket{\tilde{1}} \approx \mathcal{N}_1 \left( \frac{\Omega_2^2}{\Omega_1^2} \ket{g_1} - \frac{\sqrt{2} \Omega_2}{\Omega_1} \ket{g_2} + \ket{g_3} \right) \\
  \displaystyle \ket{\tilde{2}} \approx \mathcal{N}_2 \left( -\frac{\sqrt{2} \Omega_1^2}{\Omega_2 \Delta} \ket{g_1} - \frac{2 \Omega_1}{\Delta} \ket{g_2} - \frac{\sqrt{2} \Omega_2}{\Delta} \ket{g_3} + \frac{\Omega_1}{\Omega_2} \ket{e_1} + \ket{e_2} \right) \\
  \displaystyle \ket{\tilde{3}} \approx \mathcal{N}_3 \left( \frac{\sqrt{2} \Omega_1^2}{\Omega_2 2\Delta} \ket{g_1} + \frac{2 \Omega_1}{2\Delta} \ket{g_2} + \frac{\sqrt{2} \Omega_2}{2\Delta} \ket{g_3} + \frac{\Omega_1}{\Omega_2} \ket{e_1} + \ket{e_2} \right) \\
  \displaystyle \ket{\tilde{4}} \approx \mathcal{N}_4 \left( \frac{\sqrt{2} \Omega_2}{\Delta} \ket{g_1} - \frac{\Omega_1^2 - \Omega_2^2}{\Omega_1 \Delta} \ket{g_2} - \frac{\sqrt{2} \Omega_2}{\Delta} \ket{g_3} - \frac{\Omega_2}{\Omega_1} \ket{e_1} + \ket{e_2} \right) \\
  \displaystyle \ket{\tilde{5}} \approx \mathcal{N}_5 
   \left( -\frac{\sqrt{2} \Omega_2}{2\Delta} \ket{g_1}  + \frac{\Omega_1^2 - \Omega_2^2}{2\Omega_1 \Delta} \ket{g_2} + \frac{\sqrt{2} \Omega_2}{2\Delta} \ket{g_3} - \frac{\Omega_2}{\Omega_1} \ket{e_1} + \ket{e_2} \right)
\end{cases}
\end{equation}
\end{widetext}
where ${\cal N}_i$ are normalization factors. 
When $\Omega_2 \gg \Omega_1$, the dressed state $\ket{\tilde{1}} \approx \ket{g_1}$ which corresponds to the two molecules in the ground rovibrational level.

Until now, we have not defined the values of quantum numbers $m_1$ and $m_2$, which are selected when the polarization of the lasers is well defined, thus influencing the {}``Lambda scheme{}`` for the isolated molecule. In the following, we consider two linearly polarized beams ($q_1 = q_2 = 0$). According to Table \ref{tab:selection_rules_1photon}, this imposes $m_1 = m_2 = 0$ in the basis set in Eq. \ref{eq:bas5st}. Then DDI couples different values of $m_1$ and $m_2$ (see the next subsection). In the appendix, we have examined how the two $\pi$ lasers couple to pairs of sublevels $(m_1,m_2)$.

\subsection{R-dependent interaction Hamiltonian}

At a finite intermolecular distance $R$, when centrifugal and DDI terms come into play, all quantum numbers of the complex must be considered. Each of the five basis states of Eq.~\eqref{eq:bas5st} generates a subspace or block of Fock states with given $(N_1,N_2)$ values, and spanned by different $j_1$, $j_2$, $j_{12}$ and $\ell$ quantum numbers, since they are coupled by the DDI. The basis set becomes
\begin{equation}
\label{basis_set_5blocks}
    \begin{cases}
  |[X j_1 p_1 , X j_2 p_2],j_{12}, \ell,J,M\rangle \otimes |N_1,N_2\rangle, \\
  |[X j_1 p_1 , X j_2 p_2],j_{12}, \ell,J,M\rangle \otimes | N_1-1,N_2-1\rangle, \\
  |[X j_1 p_1 , X j_2 p_2],j_{12}, \ell,J,M\rangle \otimes | N_1-2,N_2-2\rangle, \\
  |[X j_1 p_1 , b j_2 p_2],j_{12}, \ell,J,M\rangle \otimes | N_1-1,N_2\rangle,\\
  |[X j_1 p_1 , b j_2 p_2],j_{12}, \ell,J,M\rangle \otimes | N_1-2,N_2+1\rangle,
    \end{cases}
\end{equation}
leading to the matrix of the Hamiltonian \eqref{eq:H5States-2} in a block form
\begin{widetext}
\begin{equation} \label{H_total}
\hat{H}(R) = \begin{pmatrix}
\mathbf{H_g}(R) & 0 & 0 & \mathbf{V_1} & 0 \\
 & \mathbf{H_g}(R) - \hbar \delta\omega & 0 & \mathbf{V_2} & \mathbf{V_1} \\
 & & \mathbf{H_g}(R) - 2\hbar \delta\omega & 0 & \mathbf{V_2} \\
 & & & \mathbf{H_e}(R) - \hbar \omega_{L_1} & 0 \\
 & & & & \mathbf{H_e}(R) + \hbar(\omega_{L_2} - 2\omega_{L_1})
\end{pmatrix}.
\end{equation}
\end{widetext}
The block $\mathbf{H_g}(R)$ includes the rotation of individual molecules in their $X$ state, the centrifugal term (both diagonal) and the DDI between the two molecules, to which is added the isotropic electronic van der Waals term $-C_6/R^6$. The block $\mathbf{H_e}(R)$ is similar to $\mathbf{H_g}(R)$ with one molecule in the $X$ state and the other in the $b$ state. Both blocks are themselves divided into subblocks with well-defined values of $J$ and $M$. The off-diagonal blocks $\mathbf{V_1}$ and $\mathbf{V_2}$ are the light coupling blocks due to lasers $L_1$ and $L_2$ with matrix elements calculated using Eq. \eqref{light_coupling_Rabi}. They couple blocks of $\mathbf{H_g}(R)$ and $\mathbf{H_e}(R)$ with different values of $J$.

\subsection{A first example of dressed long-range potential energy curves}
\label{ssec:example}

\begin{figure}[h!]
    \centering
    \includegraphics[scale=0.75]{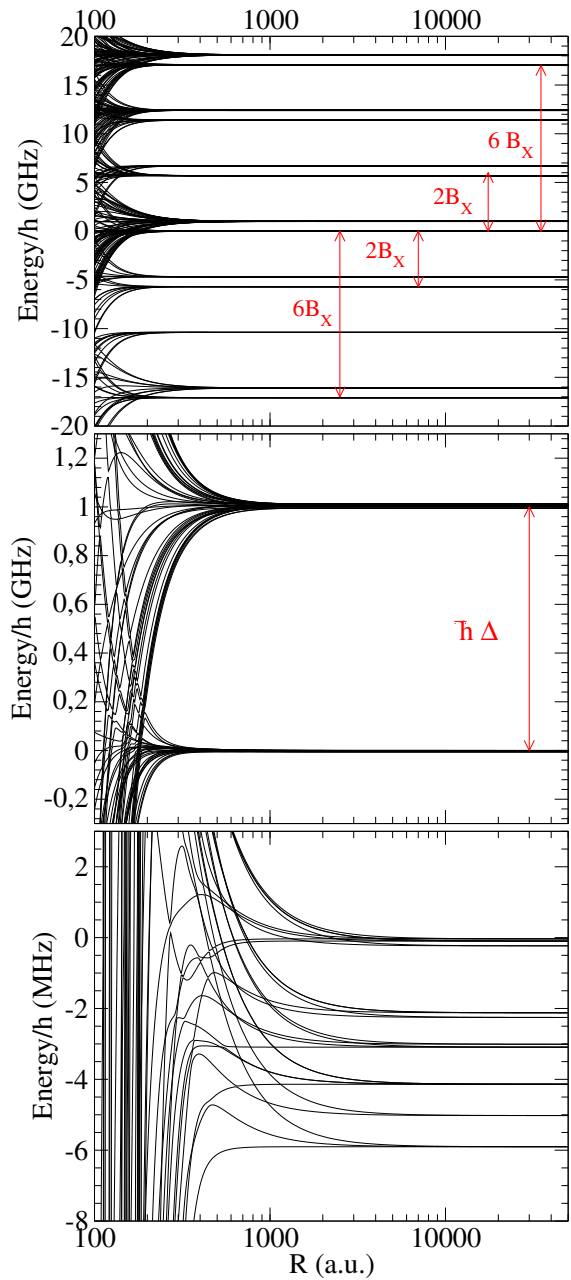}
    \caption{Dressed potential-energy curves for $\Omega_1 = 2\pi\times 50$~MHz, $\Omega_2 = 2\pi\times 200$~MHz, $\Delta = 2\pi\times 1$~GHz. Each panel corresponds to a different energy scale, with progressive zoom-ins from the upper to the lower panel.}
    \label{dressed_PECs}
\end{figure}

We diagonalize the matrix Hamiltonian (\ref{H_total}) using the basis set of Eq. \eqref{basis_set_5blocks} by varying $j_1$, $j_2$, $\ell$ and $J$ from 0 to $j_{1,{\mathrm{max}}} = j_{2,{\mathrm{max}}} = \ell_{\mathrm{max}} = 4$ and $J_{\mathrm{max}} = 12$ and with $M=0$. Figure \ref{dressed_PECs} shows the dressed potential-energy curves (PECs) from $R=100$ to $50000$ a.u.~for two linearly polarized lasers set at a Raman resonance ($\hbar \delta \omega = 6 B_X$) with $\Omega_1 = 2\pi\times 50$ MHz, $\Omega_2 = 2\pi\times 200$ MHz and $\Delta = 2\pi\times 1$ GHz. This set of light parameters will be justified later in the article. Note that following Eq. \eqref{light_coupling_Rabi}, $\Omega_\alpha= \Omega_{\alpha,1}=\Omega_{\alpha,2} $

Under realistic ultracold conditions, the two molecules in the $X$ state with $j_X=0$ collide in the $s$ wave, leading to an initial state which is even under inversion, permutation, and reflection symmetries. Thus, we write the ground-state block $\mathbf{H_g}(R)$ on the basis of the same symmetries. The excited-state block $\mathbf{H_e}(R)$, inversion symmetry is then odd with respect to inversion to fulfill selection rules (Table \ref{tab:selection_rules_1photon}), inducing parity-changing transitions. The block $\mathbf{H_e}(R)$ remains even with respect to permutation and reflection. This results in $170$ states in each ground state block and $509$ states in each excited state block. The final dimension of the Hamiltonian matrix in Eq.~\eqref{H_total}, which comprises three ground-state blocks and two excited-state blocks each in a different Fock state, is $1528 \times 1528$.

In the top panel of Fig.~\ref{dressed_PECs}, the energy range covers a few rotational manifolds of the ground state, with $B_X \approx \hbar \times 2.9$~GHz. The asymptotes are regularly separated by approximately $2B_X$, and the visible ones are $[j_X,j_X] = [0,0]$, $[1,1]$, $[2,0]$, and $[2,2]$, shifted with the energy of various photon numbers. The initial state of the collision, namely the dark state of the 5-level Hamiltonian in Eq. \eqref{eq:H5States-2} at $R \to \infty$, has an energy close to zero in the group $[0,0]$. We expect the other groups of curves, located at a few rotational splittings, to have a small influence on the dynamics of the system. Except for the one around -10~GHz (see Ref.~\cite{karam2024} for a detailed explanation), each ladder rung contains two groups of curves, separated by 1~GHz, which is equal to the detuning $\Delta$ with respect to the $X \to b$ transition.

The middle panel of Fig.~\ref{dressed_PECs} is a close-up of the upper one, covering the energy region $[0, \hbar\Delta]$. One can see two groups of PECs: one group dissociating around zero energy with eigenvectors with dominant components from the basis states of the $\mathbf{H_g}$ block (two ground-state molecules), and one group dissociating around $1$~GHz, with eigenvectors with dominant components from the basis states of the $\mathbf{H_e}$ block (one ground- and one electronically excited molecule). In the dressed-state picture, the second group is higher in energy because the lasers are red-detuned with respect to the $X\to b$ transition. Some of these PECs possess a $R^{-3}$ character due to first-order DDI, with a longer range than the other curves that are of a van der Waals type \cite{xie2020, kara2012}.

The lower panel of Fig.~\ref{dressed_PECs} displays an energy range of a few MHz around zero. Again, this group of PECs contains the one correlated to the initial state of the collision. At the Raman resonance, the states $\ket{g_1}$, $\ket{g_2}$, $\ket{g_3}$ given in Eq.~\eqref{eq:bas5st} are almost degenerate around zero energy, the degeneracy being lifted by the laser fields which couple them to $\ket{e_1}$ and $\ket{e_2}$. The number of asymptotic limits corresponds to the different values of $|m_1|$ and $|m_2|$ involved. 

\section{Adiabatic Elimination}\label{sec:AdiabElim}

On the middle panel of Fig.~\ref{dressed_PECs}, the large energy spacing between the group of PECs corresponding to two ground-state molecules and the one corresponding to one ground- and one excited molecule suggests that the latter has a small influence on the dynamics and can be treated perturbatively. This is the aim of \textit{adiabatic elimination} (AE), whose principle and validity are discussed in the present section. 

When the lasers are far off-resonant from the transition to the electronically excited state, the latter remains scarcely populated throughout the collision. This state primarily serves to mediate the coupling between the components of the electronic ground state and has a weak influence on the population dynamics.

It is therefore advantageous to derive an effective Hamiltonian in which the population of the electronic excited-state block $\mathbf{H_e}$ is eliminated, retaining only the states of the ground-state block coupled through an effective interaction. This widely used approach, known as adiabatic elimination \cite{brion2007, paulisch2014, Sinatra1995}, involves identifying a subset of states that are relevant to the evolution of the system and separating them from those that are irrelevant.

In our case, the relevant states correspond to the three ground-state blocks of the Hamiltonian (\ref{H_total}), since they include the initial state $\ket{g_1}$ and the states $\ket{g_2}$ and $\ket{g_3}$ strongly coupled to it. The irrelevant states consist of the two electronic excited-state blocks of the Hamiltonian (\ref{H_total}). By applying the method outlined in \cite{paulisch2014}, we derive an effective Hamiltonian expressed on a basis that excludes the excited electronic states,
\begin{widetext}
\begin{equation}\label{effective_hamiltonian_5level_sys}
    \hat{H}_{\text{eff}}(R) = 
    \begin{pmatrix}
        	\mathbf{H_g}(R) - \mathbf{V_1} \boldsymbol{\Delta}^{-1} \mathbf{V_1}^\dagger & -\mathbf{V_1} \boldsymbol{\Delta}^{-1} \mathbf{V_2}^\dagger & 0 \\
			& \mathbf{H_g}(R) + \hbar\boldsymbol{\delta}\omega -\mathbf{V_2} \boldsymbol{\Delta}^{-1} \mathbf{V_2}^\dagger - \mathbf{V_1} \boldsymbol{\Delta}^{'-1} \mathbf{V_1}^\dagger & -\mathbf{V_1} \boldsymbol{\Delta}^{'-1} \mathbf{V_2}^\dagger  \\
			& & \mathbf{H_g}(R) + 2\hbar\delta\omega - \mathbf{V_2} \boldsymbol{\Delta}^{'-1} \mathbf{V_2}^\dagger  \\
    \end{pmatrix},
\end{equation}    
\end{widetext}
where we defined the detuning blocks $\hbar\boldsymbol{\Delta} = \mathbf{H_e}(R\to\infty) - \hbar \omega_{L_1} $ and $\hbar\boldsymbol{\Delta'} = \mathbf{H_e}(R\to\infty) + \hbar (\omega_{L_2} - 2\omega_{L_1}) $,  and the diagonal block $\boldsymbol{\delta}\omega = (\omega_{L_1}-\omega_{L_2})\times \mathbb{I}$.

\begin{figure}[h!]
    \centering
    \includegraphics[scale=  0.7]{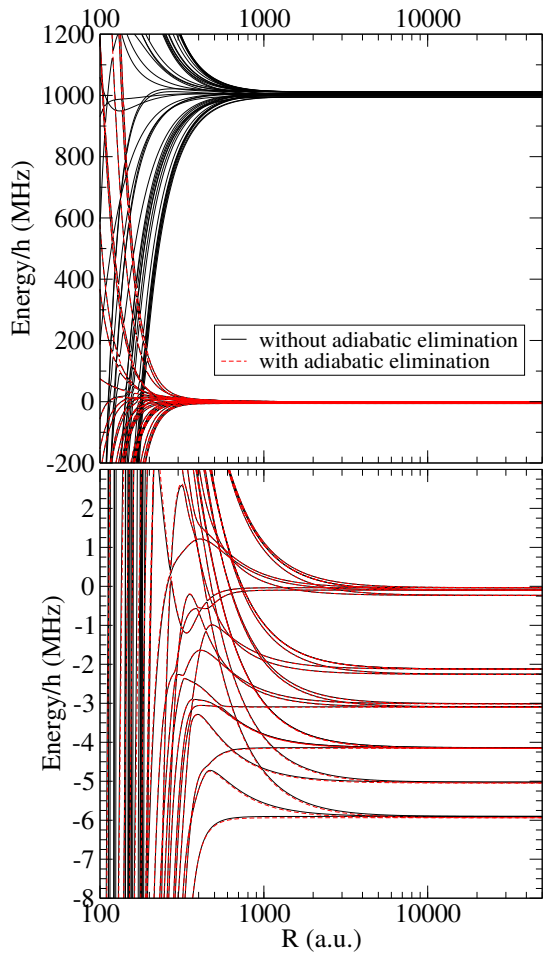}
    \caption{Adiabatic elimination: dressed PECs calculated by diagonalizing the full Hamiltonian (Eq.(\ref{H_total}), solid black lines), and the effective Hamiltonian (Eq. \ref{effective_hamiltonian_5level_sys}, in dashed red lines), for laser parameters $\Omega_1 = 2\pi\times 50$~MHz, $\Omega_2 = 2\pi\times 200$~MHz, and $\Delta = 2\pi\times 1$~GHz. The upper panel shows the PECs on a scale of 1400~MHz including the excited state manifold, and the lower panel shows a zoom-in on a scale of 11~MHz around the zero energy.
    \label{adiab_elimination_fig}
    }
\end{figure}

Figure \ref{adiab_elimination_fig} displays the dressed PECs obtained by diagonalizing the full Hamiltonian of Eq. \eqref{H_total} in solid black lines and the effective Hamiltonian of Eq.\eqref{effective_hamiltonian_5level_sys} in dashed red lines, for the same laser parameters as in Fig. \ref{dressed_PECs}. 

A key observation is the absence of dashed red lines correlated with the asymptotic limits near 1~GHz. This results from the elimination of excited states associated with Fock states $\ket{N_1, N_2} = \ket{-1,0}$ and $\ket{-2,+1}$ in the effective Hamiltonian, preventing the appearance of dressed energy levels near $\Delta = 2\pi\times 1$~GHz. However, the interaction between states in the ground-state manifold remains mediated through the eliminated excited states via the coupling terms $\mathbf{V_1}$ and $\mathbf{V_2}$, ensuring consistency with the full model.

The asymptotic limits around  zero correspond to states involving the remaining Fock states $\ket{0,0}$, $\ket{-1,+1}$, and $\ket{-2,+2}$. In the regime where $\Delta \gg \Omega_1, \Omega_2$, the weights of the excited states in the dressed states are negligible compared to those of the ground states near zero energy. We observe a very good agreement between the black and red curves, which validates AE for the chosen laser parameters.

\begin{figure}[h!]
    \centering
    \includegraphics[scale=0.7]{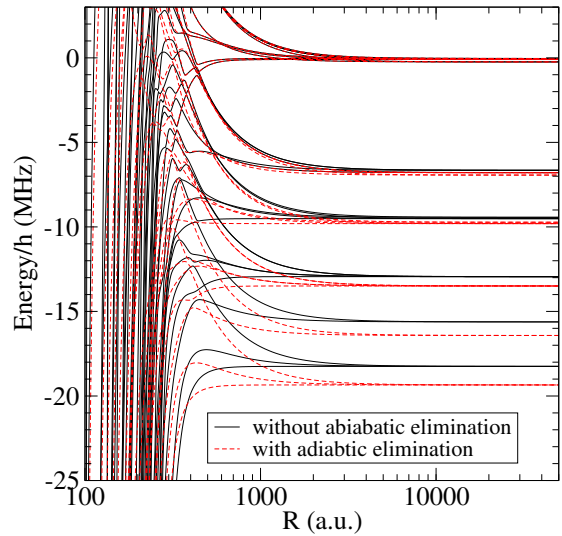}
    \caption{Invalid adiabatic elimination: same as Figure \ref{adiab_elimination_fig}, but for $\Delta = 2\pi\times 300$~MHz.
    \label{adiab_not_elim}
    }
\end{figure}

To illustrate the validity domain of AE, Fig. \ref{adiab_not_elim} shows the same PECs as in Figure \ref{adiab_elimination_fig}, but for a smaller detuning $\Delta = 2\pi\times 300$~MHz. We see that AE no longer produces PECs in agreement with the PECs derived from the full Hamiltonian. This discrepancy arises because, for fixed coupling strengths $\Omega_1$ and $\Omega_2$, the effectiveness of induced interactions increases as $\Delta$ diminishes, thus leaving the perturbative regime.

\section{Application to two-photon optical shielding of collisions}
\label{sec:2OS}

In the previous sections, we have plotted long-range PECs for various sets of light parameters, showing that intermolecular interactions can be modified to some extent. In this section, we illustrate how the laser beams can influence the collisional properties of the molecules. We use a similar method to Ref.~\cite{xie2020}, assuming that the relevant dynamics takes place in the long-range region, while molecules that sufficiently approach each other have a unit probability to undergo a reactive collision. We use the time-independent quantum-mechanical formulation of collision theory to calculate the scattering matrix, and in particular its matrix elements involving the initial state of the collision, or the entrance channel. Because we consider two molecules submitted to two fields, identifying the entrance channel is not a trivial task that should be done with caution.

\begin{figure*}[t]
\centering
\includegraphics[scale=0.65]{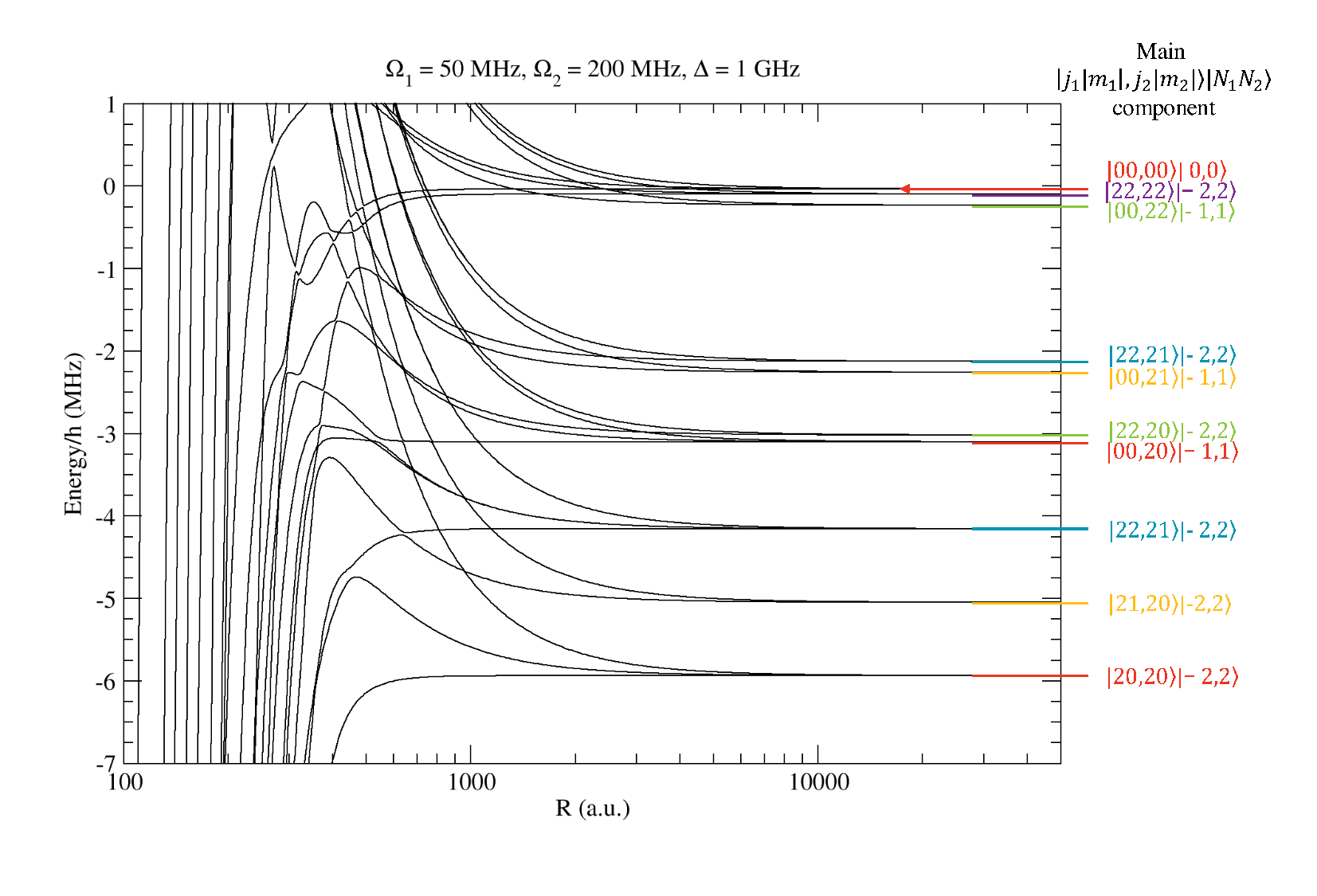}
\caption{Dressed PECs in the energy range of the entrance channel (marked with a red arrow) for $\Omega_1 = 2\pi\times 50$~MHz, $\Omega_2 = 2\pi\times 200$~MHz, and $\Delta = 2\pi\times 1$~GHz at Raman resonance ($\hbar \delta\omega = 6B_X$). The main components of each channel are listed on the right, with colors corresponding to $(|m_1|, |m_2|)$ subspaces: red for $(0,0)$, yellow for $(0,1)$, green for $(0,2)$, blue for $(1,1)$ and $(1,2)$, and purple for $(2,2)$ (see Appendix for a detailed discussion).}
\label{entrance_states_figure}
\end{figure*}

\subsection{Identifying the entrance channel}

In field-assisted collisions, the asymptotic channels $\ket{\tilde{n}}_\infty$ are no longer the bare molecular states $\ket{n}$ because of field mixing. They can be written as
\begin{equation}
  \ket{\Tilde{n}}_\infty = \sum_n c^{\infty}_{\Tilde{n},n} \ket{n}.
  \label{eq:ntilde}
\end{equation}
where in this section $\ket{n}$ is expressed in the dressed uncoupled basis (Eq.\eqref{uncoupled_basis+fock}).  

In the weak-coupling regime, the sum often includes a single dominant bare molecular state $\ket{n}$ that is important to find, in order to interpret the dressed PECs and the S-matrix elements. In this respect, Fig. \ref{entrance_states_figure} displays dressed PECs using adiabatic elimination and in a range of laser parameters for which it is possible to identify the dominant molecular states of the dissociation limits. Their labels are sorted by pairs of $(|m_1|,|m_2|)$, as different pairs are not mixed by light. The red curves are especially important, since they contain the entrance channel $\ket{j_1 = 0, m_1 = 0, j_2 = 0, m_2 = 0, \ell = 0, m_\ell = 0}$.

In the strong-coupling regime, it is no longer possible to determine a dominant state in Eq.~\eqref{eq:ntilde}. However, it is still crucial to identify the entrance channel in order to know which elements of the S-matrix should be used to calculate the rate coefficients. To do so, we focus on the curves with dissociation energies closest to zero. They contain the entrance channel, as the latter consists of two molecules in the dark state $\ket{\tilde{1}}$ of Eq.~\eqref{dressed_states_5lvl}. As Figure \ref{entrance_states_figure} shows, the {}``zero" dissociation limits also contain the molecular state $\ket{j_1=2,|m_1|=2,j_2=2,|m_2|=2}$, insensitive to lasers (see Fig.~\ref{full_level_scheme_compact}). Among these limits, the entrance channel is the only one characterized by $m_1 = m_2 = 0$ -- regardless of the weights of the different $(j_1,j_2)$ pairs --, and partial-wave quantum numbers $\ell = m_\ell = 0$. 

When $R \to \infty$, the centrifugal term vanishes, so a dark state $\ket{\Tilde{1},\ell,m_\ell}_\infty$ exists for each partial wave $(\ell,m_\ell)$, whose eigenvector is independent from the latter. Following Eq.~\eqref{dressed_states_5lvl}, we expand the dark state $\ket{\tilde{1}}$ as
\begin{equation}
  \begin{split}
  \ket{\Tilde{1},\ell,m_\ell}_\infty 
   = & c^\infty_1 \ket{00, 00, \ell,m_\ell} \otimes \ket{0, 0} \\
   + & c^\infty_2 \ket{20, 00, \ell,m_\ell} \otimes \ket{-1, +1}\\
   + & c^\infty_3 \ket{20, 20, \ell,m_\ell} \otimes \ket{-2, +2},
\end{split}
\end{equation}
where $\sum_i (c^\infty_i)^2 = 1$.

Table \ref{strength_coupling} presents the squares of the coefficients $c^\infty_i$ for various coupling parameters, ranging from no coupling ($\Omega_1 = 0$, $\Omega_2 = 2\pi\times 200$~MHz) to strong coupling ($\Omega_1 = \Omega_2 = 2\pi\times 200$~MHz). These results are obtained with adiabatic elimination and with $\Delta = 2\pi\times 1$~GHz and $\delta = 0$. As $\Omega_1$ increases, the weight of the rovibrational ground level gets smaller and smaller, until it becomes negligible.

\begin{table}[h!]
\centering
\caption{The weights $(c^\infty_1)^2$, $(c^\infty_2)^2$, $(c^\infty_3)^2$ of the asymptotic states in the entrance scattering channel, for different Rabi frequencies $\Omega_1$ at fixed $\Omega_2 = 2\pi\times 200$~MHz.}
\label{strength_coupling} 
%
\begin{tabular}{ccccc} 
 \hline \hline
  $\Omega_1/2\pi$ (MHz) & $\Omega_2/2\pi$ (MHz) &  $(c^\infty_1)^2$ & $(c^\infty_2)^2$ & $(c^\infty_3)^2$ \\ 
 \hline\hline
    0 & 200  & 1.0    & 0      & 0 \\ 
   50 & 200  & 0.880  & 0.115  & $\leq 0.05$ \\ 
  100 & 200  & 0.6134 & 0.3354 & 0.0511 \\ 
  150 & 200  & 0.3660 & 0.4692 & 0.1646 \\ 
  200 & 200  & 0.2069 & 0.4857 & 0.3073 \\ 
 \hline \hline
\end{tabular}
\end{table}

\subsection{Dynamical calculations and rate coefficients}

We propagate the log-derivative of the wavefunction of the system from $R_{\mathrm{min}}=5$~a.u. to $R_{\mathrm{max}} = 50000$~a.u. with the method of Manolopoulos \cite{manolopoulos1986} using an adaptive grid step mapping the dressed PECs \cite{kokoouline1999, willner2004}. We impose an absorbing condition at $R_{\mathrm{min}}$ modeling the experimentally observed losses from \cite{Wang2015}. The total energy of the system $E_{\mathrm{tot}} = E_{\mathrm{in}} + k_B T$ is the sum of the asymptotic energy of the entrance channel $E_{\mathrm{in}}$ increased by the collision energy $k_B T$ corresponding to a representative experimental temperature $T=300$~nK, where $k_B$ is the Boltzmann constant. After extracting the S-matrix, we calculate the elastic, inelastic, and reactive rate coefficients, $\beta_i^{\text{el}}$, $\beta_i^{\text{inel}}$, and $\beta_i^{\text{rea}}$, using 
\begin{align}
  \beta_i^{\text{el}} (E) &= g_i \frac{\pi}{\mu k_i} |1-S_{ii}(E)|^2
  \label{elastic rates} \\
  \beta_i^{\text{inel}} (E) &= g_i \frac{\pi}{\mu k_i}
    \sum_{j \neq i} |S_{ij}(E)|^2 
  \label{inelastic rates} \\
  \beta_i^{\text{rea}} (E) &= g_i \frac{\pi}{\mu k_i}
    \left( 1 - \sum_j |S_{ij}(E)|^2 \right)
  \label{reactive rates}
\end{align}
with the factor $g_i=2$ for identical colliding species and $g_i=1$ otherwise, and $k_i$ the asymptotic wave vector of the entrance channel $i$.

\begin{figure}[h!]
    \centering
    \includegraphics[scale=0.4]{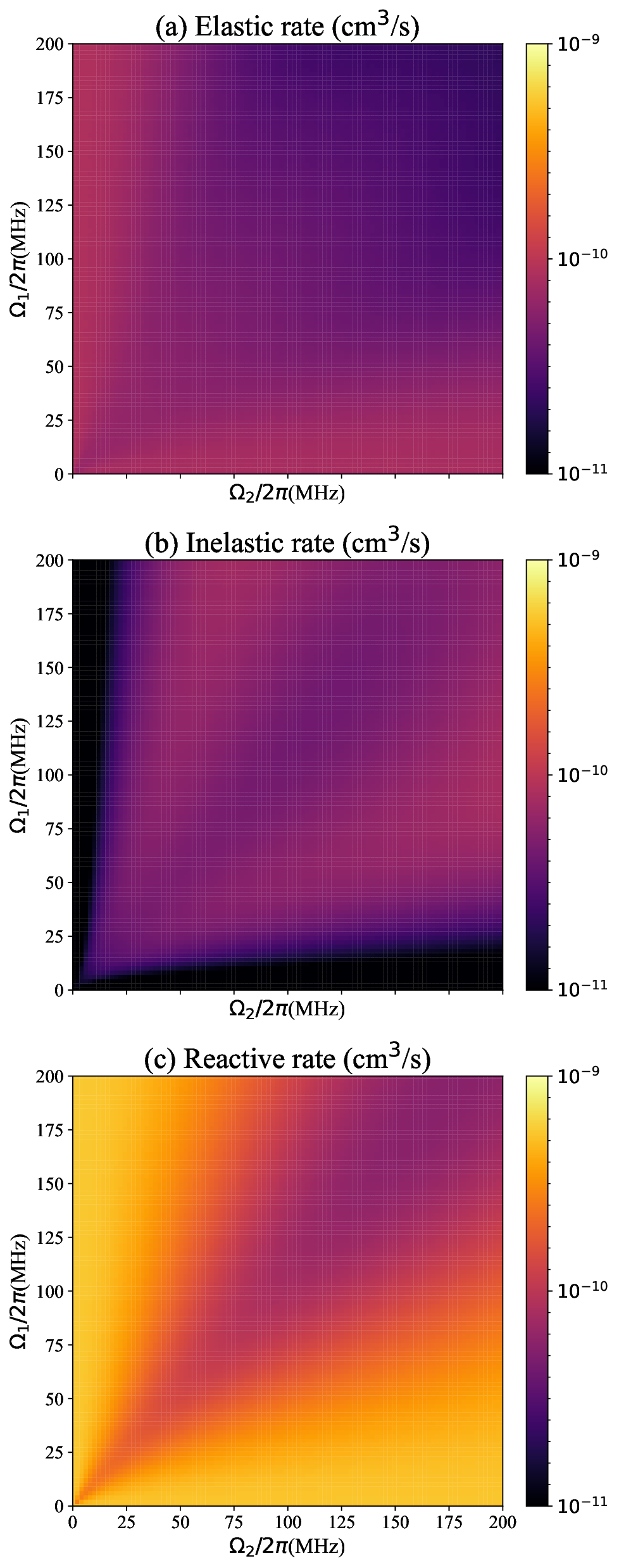}
    \caption{Collision rate coefficients (in cm$^3$/s) of two ultracold $^{23}$Na$^{39}$K molecules exposed to two lasers, as function of their Rabi frequencies $\Omega_1$ and $\Omega_2$ for a detuning $\Delta = 2\pi\times 1$~GHz, at the Raman resonance ($\delta\omega = 0$), and for a temperature of $T=300$~nK: (a) elastic, (b) inelastic and (c) reactive rates.}
    \label{fig:rates2-os}
\end{figure}

The subsequent results are obtained with adiabatic elimination using the Hamiltonian of Eq. (\ref{effective_hamiltonian_5level_sys}). 
We take a fixed detuning $\Delta = 2\pi\times 1$~GHz and the Raman resonance condition $\hbar (\omega_{L1} - \omega_{L2}) = 6 B_X$. Figure \ref{fig:rates2-os} shows the elastic, inelastic, and reactive rate coefficients as functions of the Rabi frequencies $\Omega_1$ and $\Omega_2$. The three graphs are obtained after the calculation of $10^4$ points in the $(\Omega_1,\Omega_2)$ frequency plane, with a step size of 2~MHz. Each point involves a full resolution of the coupled-channel equations and the extraction of the collision rates for the collision energy, Rabi frequencies, and detunings.
The reactive rates shown in Figure \ref{fig:rates2-os}(c) reveal important insights into the dynamics of the two-photon process. In the regime where $\Omega_1 = 0$ or $\Omega_2 = 0$, the shielding is absent and the reactive rate reaches $5.62 \times 10^{-10}$~cm$^3$/s. This value corresponds to the universal loss rate, closely matching the experimental results for \NaK molecules without shielding lasers \cite{voges2020a}. In the same regime, the inelastic rates shown in Fig. \ref{fig:rates2-os}(b) are negligible, as the entrance channel is the lowest in energy. At $\Omega_1 = \Omega_2 = 0$, the inelastic rate is exactly zero.

As Rabi frequencies increase, a notable region of reduced reactive rates emerges around the diagonal where $\Omega_1 \approx \Omega_2$. In this regime, corresponding to strong coupling, the reactive rate reaches a minimum value of $5.56 \times 10^{-11}$~cm$^3$/s for $\Omega_1 = 168$~MHz and $\Omega_2 = 186$~MHz. Although this is approximately one order of magnitude lower than the universal loss rate ($4.49 \times 10^{-10}$~cm$^3$/s \cite{voges2020a}), it remains noticeable.

In contrast, the inelastic rates of Fig. \ref{fig:rates2-os}(b) increase significantly with light coupling strength. The maximum inelastic rate occurs when one Rabi frequency is approximately half of the other, exceeding the elastic rate. To understand this result, we examine S-matrix elements related to the open exit channels identified in Fig. \ref{entrance_states_figure}. Specifically, we calculate the transition probability from the entrance channel to all open channels, quantified by $|S_{ij}|^2$, where $i$ represents the entrance channel and $j$ an exit channel. We consider the $s$-wave entrance dark state as the initial channel.

\begin{table}[h!]
\begin{center}
\caption{Asymptotic dressed energies of the entrance and open exit channels with their main component and weight in the dressed adiabatic state as well as the branching ratio giving the percentage of a wavefunction to end up in a given exit channel.}
\label{S_matrix_table} 
\begin{tabular}{cccc} 
 \hline
 \hline
Energy& Main component & \multirow{2}{*}{Weight} & Branching\\ [0.2ex] 
 (MHz)  & $\ket{j_1 |m_1|, j_2 |m_2|} \ket{N_1,N_2}$&   & ratio\\[0.25ex] 
 \hline\hline 
-0.0349 &$\ket{00,00} \ket{0,0}$ & 0.880 & -\\[0.25ex]
 \hline 
-0.0968 &$\ket{22,22} \ket{-2,2}$ &0.999 &  0\%\\[0.25ex]
\hline 
-0.2320 &$\ket{00,22} \ket{-1,1}$ &0.922 &  0.006 \% \\[0.25ex]
\hline 
-2.125 &$\ket{22,21} \ket{-2,2}$ & 0.815 & 0\%\\[0.25ex]
\hline
-2.257 &$\ket{00,21} \ket{-1,1}$ & 0.922&0.56\%\\[0.25ex]
\hline
-3.017 & $\ket{22,20} \ket{-2,2}$ & 0.922  & 0.16\% \\[0.25ex]
\hline
-3.100 & $\ket{00,20} \ket{-1,1}$ & 0.742  & 88.32\%\\[0.25ex]
\hline
-4.154 & $\ket{21,21} \ket{-2,2}$ & 0.735 & 0.37\% \\[0.25ex]
\hline
-5.046 & $\ket{21,20} \ket{-2,2}$&0.922  & 0.50 \%\\[0.25ex]
\hline
-5.9359 & $\ket{20,20} \ket{-2,2}$& 0.8532 & 10.06\% \\[0.25ex]
\hline \hline
\end{tabular}
\end{center}
\end{table}

Table \ref{S_matrix_table} presents the asymptotic dressed energies of the entrance and open channels, their dominant components and weights in the dressed state, and the probability of ending in the open channel. The main portion of the initial wavefunction ends after the collision in the open channels at asymptotic energies $-3.100$~MHz and $-5.9359$~MHz, represented by the red lines in Figure \ref{entrance_states_figure}.

These channels correspond to the states $\ket{g_2}$ and $\ket{g_3}$ of Eq.~\eqref{eq:bas5st} with $m_1 = m_2 = 0$, due to the $\pi$ polarization of the beams. Smaller probability are observed toward other channels with $m_1$ or $m_2 \neq 0$ which are not coupled to the entrance channel by the lasers, but by the second-order dipole-dipole (van der Waals) interaction (see Section \ref{ssec:example}).

\section{Conclusions and prospects}
\label{sec:ccl}

In this article, we developed the quantum formalism to treat collisions of ground-state ultracold molecules submitted to two cw lasers. We assume that the collisional dynamics takes place essentially at large intermolecular distances, where the dipole-dipole and van der Waals interactions are dominant, while full losses due to chemical interactions are assumed at smaller distances. The rate coefficients of elastic, inelastic, and reactive (loss) collisions are calculated using the time-independent quantum formalism based on Manolopoulos' propagator of the wave function log-derivative. 

The field-molecules Hamiltonian is described in the dressed, symmetrized fully-coupled basis, including the photon numbers in both laser feilds. The isolated molecules are initially modeled by a 3-level {}``Lambda'' scheme, which involves an excited electronic state in the upper level. In the asymptotic region, the lasers-molecules system is described with a five-level scheme equivalent to the superposition of two identical 3-level {}``Lambda'' scheme. At finite distance, the interaction between the molecules generates a dynamical ($R$-dependent) five-level system, expressed as a five-block Hamiltonian. We thus demonstrate that the two lasers modify the long-range interactions between the molecules and thus their dynamics.

When laser frequencies are tuned at the Raman resonance, individual molecules are prepared in a dark state independent of the electronic excited state, and in which the photon scattering and spontaneous emission are suppressed, thus protecting the ultracold molecular sample from unwanted heating. When the laser frequencies are detuned to the red from the molecular resonance by a large amount, the five-block Hamiltonian can be reduced to a three-block Hamiltonian using adiabatic elimination of the excited state, reducing the basis size and so the computational time.  We check the validity of adiabatic elimination in conditions where the influence of the electronic excited state can be treated in a perturbative way.

With our formalism, we explored the feasibility of a two-photon optical scheme to suppress ultracold  molecular collisions at short intermolecular distances, the so-called two-photon optical shielding (2-OS) \cite{karam2023}. In the explored range of laser parameters, we observed significant variations of both the inelastic and the reactive rates. However, we did not identify clear shielding conditions in which the elastic rate induced by the two-photon process consistently surpasses both the inelastic and reactive rates. 

As optical photons represent exquisite control knobs in an experiment, it is worth pursuing an in-depth study of 2-OS with the present formalism. First, the dependence of the scattering rates on the detuning $\Delta$ of the single-molecule {}``Lambda'' scheme could be investigated, for example by reducing the red detuning $\Delta$ beyond the validity of adiabatic elimination, and even considering blue detunings, so that the excited state could bring an additional control parameter to 2-OS. Moreover, as a static electric field is applied in the ongoing experiment to induce the permanent electric dipole moment of the molecule in the SF frame, the present formalism could be generalized by including such a field in the Hamiltonian, which will presumably enforce the DDI between the molecules during their collision.

\section{Acknowledgments}
C. K. acknowledges the support of the Quantum Institute of Université Paris-Saclay. Stimulating discussions with Prof. Eberhard Tiemann, Prof. Silke Ospelkaus, and Dr Leon Karpa (IQO, Leibniz University, Hannover) are gratefully acknowledged. This work is supported by the grant ANR-22-CE92-0069-01 (OPENMINT project) from the Agence Nationale de la Recherche.

\section*{Appendices}

\begin{figure*}[h!]
    \centering
    \includegraphics[scale=0.45]{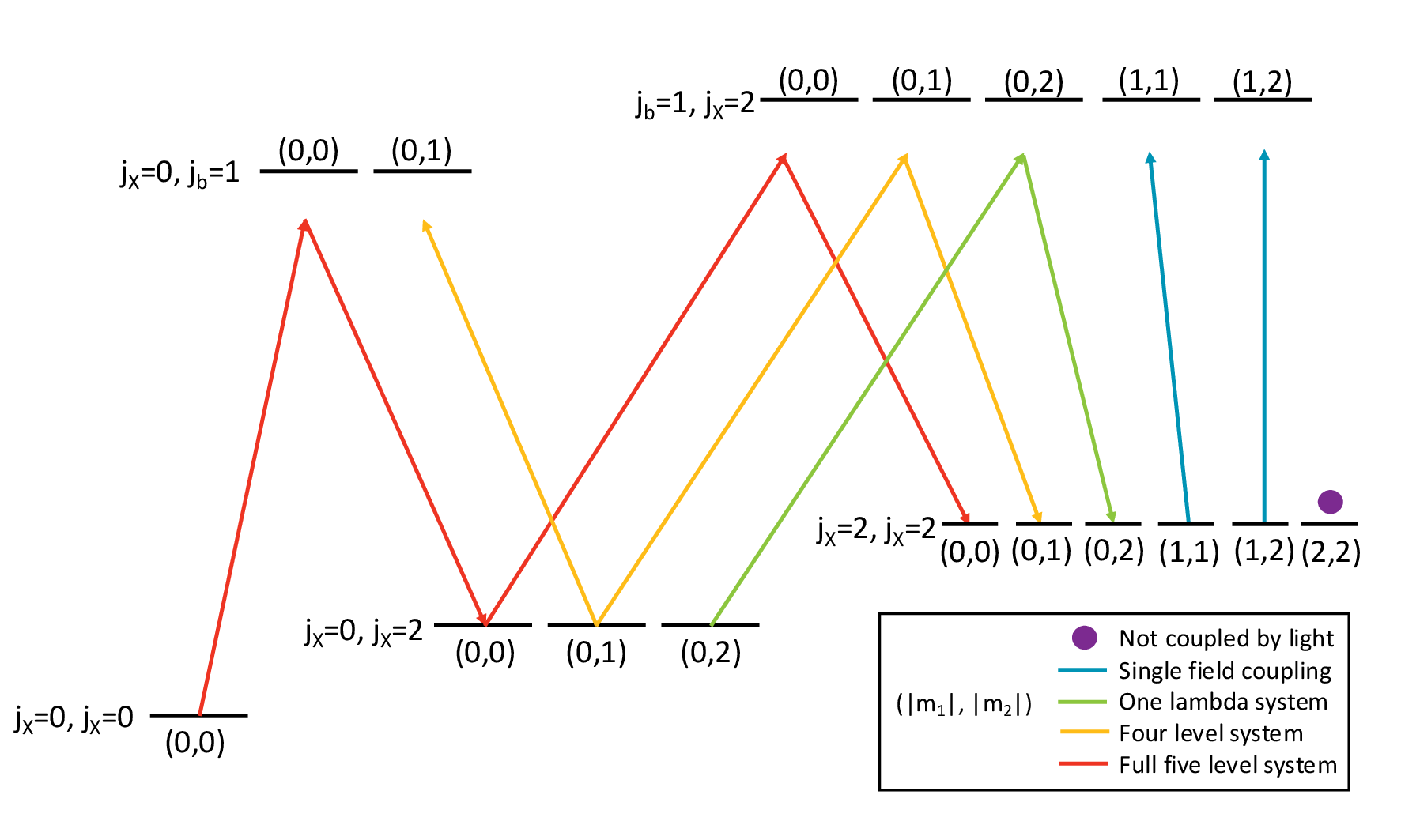}
    \caption{The coupling scheme of two molecules exposed to two linearly polarized $\pi$ laser fields. Solid red arrows represent the five-level coupling scheme in the $(|m_1|,|m_2|)=(0,0)$ subspace. Solid yellow arrows represent the four-level scheme in the $(|m_1|,|m_2|)=(0,1)$ subspace. Solid green arrows represent the three-level system, and a single solid yellow arrow represents a two-level scheme in the $(|m_1|,|m_2|)=(0,2)$ subspace. Solid blue arrows represent the two-level schemes in the $(|m_1|,|m_2|)=(1,1)$ and $(|m_1|,|m_2|)=(1,2)$ subspaces. The purple dot corresponds to the state that is not coupled to lasers in the $(|m_1|,|m_2|)=(2,2)$ subspace.}
    \label{full_level_scheme_compact}
\end{figure*}

The projections $m_i$ of the angular momentum $j_i$ are central to understanding the subspace structures for non-interacting molecules, as the asymptotic conditions of our formalism. We consider here all allowed projections, $|m_i|<j_i$ and restrict the study to two linearly polarized lasers. 
Our two-photon scheme originating from $\ket{g_1}$ includes subspaces characterized by $|m_1| + |m_2| = 0, 1, \dots, 4$. Table \ref{Five_level_system_table} lists the basis states for each subspace, while Figure \ref{full_level_scheme_compact} illustrates the coupling schemes. These schemes connect the rotational states $(j_1, j_2)$ of the molecular pair and their components, initially degenerate in energy, before laser couplings are applied.

\begin{figure*}[p]
    \centering
    \includegraphics[scale=0.9]{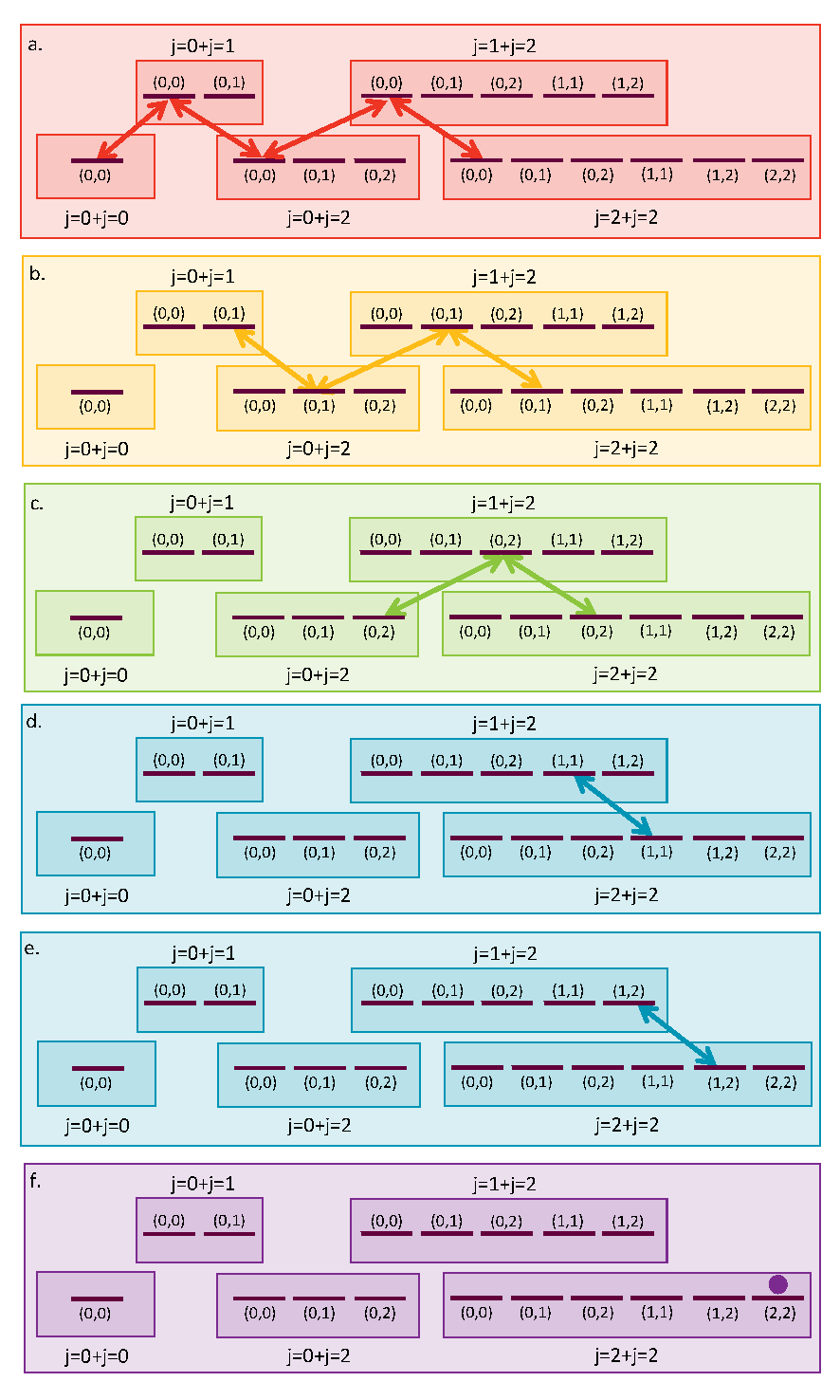}
    \caption{The six coupling schemes for the different $(|m_1|+|m_2|)$ values in the rotating wave frame in the case of two linearly polarized lasers ($\Delta m_1=0$, $\Delta m_2=0$). The inner five rectangles of each panel correspond to the different $j_1+j_2$ values. The arrows represent the laser couplings between the different $(|m_1|,|m_2|)$ levels. Panels a, b, c, d, e, and f represent the five-, four-, three-level systems, single field coupling, and the uncoupled state, respectively.} 
    \label{full_level_scheme}
\end{figure*}

Qualitatively, it is interesting to distinguish coupling schemes corresponding to different $(|m_1|, |m_2|)$ subspaces.
\begin{itemize}
    \item Subspace $(|m_1|, |m_2|) = (0,0)$: The Hamiltonian eigenstates include a dark state, $\ket{\tilde{1}}$, insensitive to laser-induced energy shifts. The coupling forms a five-level system, as shown in Figure \ref{full_level_scheme}.a.
    
    \item Subspace $(0,1)$: Despite having five states, this subspace functions as a four-level system because two excited states share the same Fock state $\ket{-2,+1}$, resulting in four laser transitions (Figure \ref{full_level_scheme}.b).
    
    \item Subspace $ (0,2)$: This three-level $\Lambda$ system supports a dark state formed as a linear combination of $\ket{g, 0, \mathbf{0}, +1, G, 2, \mathbf{2}, +1}$ and $\ket{g, 2, \mathbf{0}, +1, g, 2, \mathbf{2}, +1}$, excluding the absolute ground state (Figure \ref{full_level_scheme}.c).
    
    \item Subspaces $ (1,1)$ and $(|m_1|, |m_2|) = (1,2)$: These are single-photon couplings, with two-level Rabi oscillations driven by $\Omega_2$ (Figures \ref{full_level_scheme}.d and \ref{full_level_scheme}.e).
    
    \item Subspace $(2,2)$: This uncoupled state is unaffected by light due to the selection rules for $\pi-\pi$ transitions. Its energy remains unchanged, but does not form a dark state (Figure \ref{full_level_scheme}.f).
\end{itemize}

The different schemes are identified by the condition $\mathbf{j_i \geq |m_i|}$ that depends on the value of $m_i$ and not its sign. We can thus establish a table similar to Table (\ref{Five_level_system_table}), but restricted to ($-|m_1|, |m_2|$), ($|m_1|, -|m_2|$) and ($-|m_1|, -|m_2|$). The states are all degenerate with the ones listed for ($|m_1|, |m_2|$) in Table \ref{Five_level_system_table}.

\begin{table*}[h!]
\begin{center}
\caption{Basis set of different $(|m_1|,|m_2|)$ sub-spaces for a $\pi-\pi$ polarized two-photon transition. For each subspace we give the involved values of the individual projections, followed by the symmetrized wavefunction and the Fock state. For each subspace we define a coupling scheme illustrated in the figures quoted in the last column.}
\label{Five_level_system_table} \begin{tabular}{ccccc} 
 \hline
 \hline
 ($|m_1|,|m_2|$)& $\ket{e_1,j_1,|m_1|,p_1,e_2,j_2,|m_2|,p_2}_+$ & $\ket{N_1, N_2}$& Coupling scheme& Figure\\ [0.5ex] 
 \hline
\multirow{5}{*}{($0,0$)} & $\ket{X,0,\mathbf{0},+1,X,0,\mathbf{0},+1}$& $\ket{0, 0}$& \multirow{5}{*}{Five level system}& \multirow{5}{*}{\ref{full_level_scheme}.a}\\
 & $\ket{X,0,\mathbf{0},+1,X,2,\mathbf{0},+1}$& $\ket{-1, +1}$& & \\
 & $\ket{X,2,\mathbf{0},+1,X,2,\mathbf{0},+1}$& $\ket{-2, +2}$& & \\[0.5ex]
 & $\ket{X,0,\mathbf{0},+1,,b,1,\mathbf{0},-1}$& $\ket{-1, 0}$& & \\
 & $\ket{X,2,\mathbf{0},+1,b,1,\mathbf{0},-1}$& $\ket{-2, +1}$& &\\ [0.25ex]
 \hline
 \multirow{5}{*}{($0,1$)} & $\ket{X,0,\mathbf{0},+1,X,2,\mathbf{1},+1}$& $\ket{-1, +1}$&\multirow{5}{*}{Four level system}&\multirow{5}{*}{\ref{full_level_scheme}.b}\\ 
 & $\ket{X,2,\mathbf{1},+1,X,2,\mathbf{0},+1}$& $\ket{-2, +2}$& & \\[0.5ex]
 & $\ket{X,0,\mathbf{0},+1,b,1,\mathbf{1},-1}$& $\ket{-1, 0}$& & \\
 & $\ket{X,2,\mathbf{1},+1,b,1,\mathbf{0},-1}$& $\ket{-2, +1}$& & \\
 & $\ket{X,2,\mathbf{0},+1,b,1,\mathbf{1},-1}$& $\ket{-2, +1}$& & \\[0.25ex]
 \hline
 \multirow{3}{*}{($0,2$)} & $\ket{X,0,\mathbf{0},+1,X,2,\mathbf{2},+1}$& $\ket{-1, +1}$& \multirow{3}{*}{Three level $\Lambda$ system}&\multirow{3}{*}{\ref{full_level_scheme}.c}\\ 
 & $\ket{X,2,\mathbf{2},+1,X,2,\mathbf{0},+1}$& $\ket{-2, +2}$& & \\[0.5ex]
 & $\ket{X,2,\mathbf{2},+1,b,1,\mathbf{0},-1}$& $\ket{-2, +1}$& & \\
 \hline
\multirow{2}{*}{($1,1$)} & $\ket{X,2,\mathbf{1},+1,X,2,\mathbf{1},+1}$& $\ket{-2, +2}$& \multirow{2}{*}{Single field coupling}& \multirow{2}{*}{\ref{full_level_scheme}.d} \\[0.5ex]
 & $\ket{X,2,\mathbf{1},+1,b,1,\mathbf{1},-1}$& $\ket{-2, +1}$& & \\[0.25ex]
 \hline
\multirow{2}{*}{($1,2$)} & $\ket{X,2,\mathbf{1},+1,X,2,\mathbf{2},+1}$& $\ket{-2, +2}$&\multirow{2}{*}{Single field coupling}&\multirow{2}{*}{\ref{full_level_scheme}.e}\\ 
 & $\ket{X,2,\mathbf{1},+1,b,1,\mathbf{1},-1}$& $\ket{-2, +1}$& & \\[0.25ex]
 \hline
 ($2,2$)& $\ket{X,2,\mathbf{2},+1,X,2,\mathbf{2},+1}$& $\ket{-2, +2}$&No coupling& \ref{full_level_scheme}.f\\ 
 \hline
 \hline
\end{tabular}
\end{center}
\end{table*}


\bibliographystyle{ieeetr}

\end{document}